\documentclass[prx,twocolumn,amsmath,amssymb,floatfix]{revtex4-2}
\usepackage{graphicx, amsfonts, microtype, mathrsfs, bbm, mathtools, enumerate, braket, algorithm, algpseudocodex, booktabs,colortbl}
\definecolor{myblue}{RGB}{95, 141, 211}
\usepackage[colorlinks,allcolors=myblue]{hyperref}
\usepackage[capitalize]{cleveref}

\newtheorem{definition}{Definition}

\DeclareMathOperator{\tr}{tr}
\DeclareMathOperator{\poly}{poly}
\DeclareMathOperator{\supp}{supp}
\newcommand{\1}{\mathbbm{1}}
\newcommand{\norm}[1]{\lVert #1 \rVert}
\renewcommand{\O}{\mathcal O}

\newcommand{\BQP}{\textsf{BQP}}

\renewcommand{\P}{\mathcal P}
\newcommand{\V}{\mathcal V}

\newcommand{\E}{\mathcal E}
\newcommand{\eps}{\epsilon}
\renewcommand{\H}{\mathcal H}
\newcommand{\A}{\mathcal A}
\newcommand{\B}{\mathcal B}
\newcommand{\C}{\mathbbm C}
\newcommand{\dagg}{^\dagger}

\usepackage{tikz,ifthen}
\usetikzlibrary{decorations.pathreplacing,calligraphy,decorations.markings,math}

\colorlet{tensor}{myblue}
\definecolor{ancilla}{rgb}{0.9,0.3,.3}
\definecolor{swap}{rgb}{0.9,.9,.6}

\begin{document}
\title{Computational complexity of isometric tensor network states}
\author{Daniel Malz}
\affiliation{Department of Mathematical Sciences, University of Copenhagen, 2100 Copenhagen, Denmark}
\author{Rahul Trivedi}
\affiliation{Electrical and Computer Engineering, University of Washington, Seattle, Washington 98195, USA}

\begin{abstract}
	We determine the computational power of isometric tensor network states (isoTNS), a variational ansatz originally developed to numerically find and compute properties of gapped ground states and topological states in two dimensions.
	By mapping 2D isoTNS to 1+1D unitary quantum circuits, we find that computing local expectation values in isoTNS (including those with translation-invariant bulk) is \BQP-complete.
	We then introduce \emph{injective} isoTNS, which are those isoTNS that are the unique ground states of frustration-free Hamiltonians, and which are characterized by an injectivity parameter $\delta\in(0,1/D]$, where $D$ is the bond dimension of the isoTNS.
	We show that injectivity necessarily adds depolarizing noise to the circuit at a rate  $\eta=\delta^2D^2$.
	We show that weakly injective isoTNS (small $\delta$) are still \BQP-complete, but that there exists an efficient classical algorithm to compute local expectation values in strongly injective isoTNS ($\eta\geq0.41$).
	We also show that while weakly injective isoTNS can support long-range correlations, strongly injective isoTNS are approximate Markov states with exponentially decaying correlations.
	Sampling from isoTNS corresponds to monitored quantum dynamics and we exhibit a family of isoTNS that undergo a phase transition from a hard regime to an easy phase where the monitored circuit can be sampled efficiently.
	Our results can be used to design provable algorithms to contract isoTNS.
	Our mapping between ground states of certain frustration-free Hamiltonians to open circuit dynamics in one dimension fewer may be of independent interest.
\end{abstract}
\maketitle

\section{Introduction}
One of the central goals in condensed matter theory and many-body physics is to describe and analyze ground states of local Hamiltonians.
Since the dimension of the Hilbert space of a lattice system grows exponentially with the number of sites, computing properties of even modestly sized systems requires approximations.
One particular successful strategy has been to variationally search for ground states in restricted families of states.
For example, the celebrated density matrix renormalization group method~\cite{White1992} for one dimensional systems is effectively a variational method over the class of matrix product states (MPS)~\cite{Ostlund1995,Schollwoeck2011}.
MPS are so successful, because they efficiently capture ground states of gapped Hamiltonians~\cite{Verstraete2006}, and simultaneously are easy to use on classical computers (in terms of computational complexity)~\cite{Perez-Garcia2007}.
In recent years, it has also become increasingly relevant that MPS can also efficiently be prepared on quantum computers using a linear depth circuit~\cite{Schoen2005}, or approximately even in logarithmic depth~\cite{Malz2023}.

In two and higher dimensions, the MPS ansatz has been extended to projected entangled pair states (PEPS)~\cite{Hieida1999,Maeshima2001,Verstraete2004}, which are also highly expressive~\cite{Verstraete2006a} and have been used successfully to gain insight into long-standing problems in condensed-matter physics~\cite{Jordan2008,Corboz2014,Corboz2014a,Liao2017,Zheng2017,Chen2018a,Jimenez2021,Liu2022a}.
However, the usefulness of PEPS for numerical calculations limited by the fact that contracting them in general is computationally hard (\textsf{\#P}-complete~\cite{Schuch2007}), and thus even using a quantum computer, they are hard to prepare~\cite{Schuch2007} or contract~\cite{Arad2010}.

While approximate contraction strategies for PEPS have brought considerable success~\cite{Jordan2008,Corboz2014,Corboz2014a,Liao2017,Zheng2017,Chen2018a,Jimenez2021,Liu2022a}, another strategy is to define subsets of PEPS that admit efficient algorithms to optimize them. 
An early example of such a simplification were sequentially generated states~\cite{Banuls2008}.
Leveraging MPS techniques, these states allow for efficient computation of expectation values everywhere, but unlike PEPS, increasing bond dimension seems to improve their expressibility only modestly~\cite{Banuls2008}.
More recently, isometric tensor network states (isoTNS) have been introduced~\cite{Zaletel2020}, which generalize MPS in a less restrictive way.
For instance, they provably capture exactly the renormalization-group fixed point of all string-net liquids~\cite{Soejima2020}, a constant-depth quantum circuits~\cite{Soejima2020}, some thermal states~\cite{Kadow2023} and certain fine-tuned critical states~\cite{Liu2024} and to some degree also quantum dynamics~\cite{Lin2022,Kim2023}.

IsoTNS may inhabit a sweet spot, being computationally strictly easier than PEPS, yet with some evidence towards their success.
To understand better what certain ans\"atze can and cannot capture, it is important to understand the classical and computational complexity associated to using them.
This has motivated work to characterize the (quantum and classical) computational complexity of PEPS~\cite{Schuch2007} and injective PEPS~\cite{Anshu2024}.
Here, we therefore explore the computational complexity of isoTNS. Our results hold equally well for plaquette-PEPS~\cite{Wei2022}.

\setlength{\tabcolsep}{8pt}
\begin{table*}[t]
\caption{\label{tab:comparison}
    Summary of results of this paper.
    We determine the classical complexity of computing local expectation values to constant additive error, producing samples according to the probability distribution obtained by measuring the state to constant additive error, and producing samples to multiplicative error. 
    For a specific family of injective isoTNS we can show that sampling undergoes some sort of ``measurement-induced phase transition'' to an easy phase, but we do not know what happens for general strongly injective isoTNS, hence the question mark.
}
\begin{ruledtabular}
\begin{tabular}{lcccc}
	&
	\textrm{Weakly/Non injective isoTNS}&
	\textrm{Strongly injective isoTNS}\\
	\midrule
	\textsc{expectation to $O(1)$ additive error}  & Hard unless \textsf{BQP = BPP} & Easy\\
	\textsc{sampling to $O(1)$ additive error} & Hard unless \textsf{BQP = BPP} & ?\\
	\textsc{sampling to $O(1)$ multiplicative error}  & Hard unless \textsf{postBQP = postBPP} & Hard unless \textsf{postBQP = postBPP}\\
\end{tabular}
\end{ruledtabular}
\end{table*}

\section{Summary of results}
We build on the result that isoTNS can be written in terms of a linear-depth sequential quantum circuit and thus can be regarded as being generated from a chain of ancillas that sequentially interact with the physical sites of the lattice~\cite{Wei2022}.
The time evolution of the ancillas after tracing the physical sites maps the 2D isoTNS to a 1+1D circuit of quantum channels, which needs to be simulated to compute a local expectation value on a site corresponding to some later time, which provides a way to efficiently compute expectation values in isoTNS using a quantum computer.
Our first result in \cref{sec:worst_case_hardness} is then to construct isoTNS where computing local expectation values is \BQP-complete~\cite{NielsenAndChuang}, which means that a machine or algorithm that can compute local expectation values in isoTNS is as powerful as a quantum computer.

Mirroring injective PEPS, which are those states that are unique ground states of frustration-free parent Hamiltonians)~\cite{Perez-Garcia2008}, we introduce injective isoTNS.
They are characterized by an injectivity parameter $\delta\in(0,1/D]$ which is defined as the size of the smallest singular value of the PEPS tensor.
In \cref{sec:injective_isoTNS}, we show that injectivity necessarily adds depolarizing noise at a rate  $\eta=\delta^2D^2$ to the 1+1D circuit of the ancillas, where $D$ is the bond dimension of the isoTNS.
Our second result in \cref{sec:hard_regime} is then that local expectation values in weakly injective isoTNS is still \BQP-complete, which we show by embedding a fault-tolerant circuit that is robust to small amounts of noise.
Conversely, in \cref{sec:complexity_transition}, we provide an explicit algorithm to compute local expectation values that runs in $\poly(1/\eps)$ time (independent of system size) for strongly injective isoTNS for which $\eta\geq0.41$.

We can use this results to establish two interesting physical properties of injective isoTNS, depending on their injectivity parameter (\cref{sec:physical-properties}). On the one hand, we show that weakly injective isoTNS support long-range correlations. On the other hand, strongly injective isoTNS become approximate Markov states in which the mutual information between two regions (and thus also correlations) decays exponentially with distance.

Sampling from states to additive error is at least as hard as computing local observables, so the hardness of local expectation values extends to sampling (\cref{sec:hard_sampling}).
However, we show that there are families of isoTNS in which sampling maps to monitored quantum circuits, which are known to become easy to simulate under strong monitoring~\cite{Skinner2019}.
This allows us to construct a class of isoTNS that as a function of the injectivity parameter also undergoes such a complexity transition (\cref{sec:MIPT}), which provides an intriguing link between 2D states and 1D dynamics.
We also consider sampling to multiplicative precision, which we find is hard even in the maximally injective case.

We summarize our results in \cref{tab:comparison}.

\section{Related prior work}
Schuch et al.~\cite{Schuch2007} have used a duality between PEPS and postselected quantum circuits to show that a machine capable of preparing any PEPS is equivalent to a postselected quantum computer. 
In contrast, by virtue of the isometry condition, isoTNS are dual to quantum circuits, which naturally lead to the result that a machine that can prepare isoTNS is equivalent to a quantum computer.

Anshu et al.~\cite{Anshu2024} use a similar (to the present work) fault-tolerant embedding of quantum circuits in injective PEPS to show that injective PEPS are \BQP-hard. They also show that determining expectation values of non-local observables to multiplicative precision is \#P-hard.
In injective isoTNS, we find that local expectation values are \BQP-complete and thus hard on classical computers.
Note that while it is clear that isoTNS can be prepared on a quantum computer, this is not clear for injective PEPS as considered in Ref.~\cite{Anshu2024}; it may well be the case that their complexity is greater than \BQP.
Our results establishing that computing local expectation values in sufficiently injective isoTNS is classically easy draw on early results showing that noisy quantum computers undergo a complexity transition~\cite{Aharonov1996,Aharonov2000}, which uses a similar percolation result to the present paper.

We further exhibit a family of isoTNS that (provably) undergoes a complexity transition between a regime in which approximate sampling to additive precision is classically hard, to one in which it is classically easy.
The hardness in this construction comes again from embedding a fault-tolerant circuit on the virtual degrees of freedom, whereas the easy phase arises, because in our construction certain measurement outcomes break the entanglement with a known probability, which allows us to use a percolation argument again, similar to those used to establish easiness in monitored circuits~\cite{Skinner2019,Chan2019,Li2019b}.

Many results exist on the hardness of sampling~\cite{Terhal2002,Hangleiter2023}.
To show that sampling from isoTNS to multiplicative error is hard independent of the injectivity parameter, we establish a link between maximally injective isoTNS and IQP circuits (circuits of commuting gates), which are known to be hard to sample from unless \textsf{postBQP}$=$\textsf{postBPP}~\cite{Bremner2010}.
The connection between sampling from states generated from constant-depth circuits and monitored dynamics in one dimension fewer has been expressed in Refs~\cite{Choi2019,Lu2021a,Napp2022,Bao2024}, and conceptually originates from the cluster state~\cite{Briegel2001}.
However, most recent works in monitored circuits or sampling from constant-depth circuits consider random problem instances.
Technically, hardness of approximate sampling in these settings remains a conjecture~\cite{Hangleiter2023}. Since the isoTNS we consider are neither constant-depth circuits nor random, we can prove a much stronger complexity-theoretic result.

Another result that establishes conditions under which there is an efficient classical algorithm to compute local expectation values in injective PEPS is obtained in Ref.~\cite{Schwarz2017}.
There, it was shown that local observables in a normalized injective PEPS on $d-$dimensions can be computed to a precision $\eps$ by just restricting the PEPS to a $\log(\eps^{-1})$ length cube around the local observable.
In $2D$, this would also yield a $\poly(\eps^{-1})$ algorithm for computing the local observable.
The key distinction between this and our result is that the result in Ref.~\cite{Schwarz2017} does not rely on isometry,
but instead they require that the frustration-free parent Hamiltonian of the PEPS is gapped and remains uniformly gapped when removing terms, which is an unrelated and potentially stronger condition.

IsoTNS can also be viewed as being prepared from a chain of ancillas interacting sequentially with arrays of qubits~\cite{Wei2022}, similar to how MPS can be prepared from a single emitter~\cite{Schoen2005}.
This links efforts to find efficient classical numerical approaches based on tensor networks to efforts to utilize qubit-efficient ``holographic'' ways of simulating quantum systems~\cite{Osborne2010,Kim2017,Kim2018,Borregaard2021,Niu2022,Anand2023}.
Our work directly applies to states generated in such schemes and may thus also be useful to understand the power and limitation of such approaches.
Indeed, our work shows that the ability of isoTNS to represent long-range correlations depends on injectivity parameter, with direct consequences for their expressivity.
In this context, previous research has shown that isoTNS may represent topologically ordered states~\cite{Soejima2020} and even exhibit long-range correlations~\cite{Liu2024}, it has also been shown that random isoTNS are short-range correlated~\cite{Haag2023}.

\section{isoTNS, quantum circuits, and channel circuits}
Most of our results are based on interpreting isoTNS contraction as physical evolution of the virtual degrees of freedom, which builds on representing them as sequential circuits~\cite{Wei2022}, which we recall in the following.

\subsection{Matrix product states (MPS)}
Recall that MPS of bond dimension $D$ and physical dimension $d$ can be thought of as being generated sequentially by a $D$-dimensional ancilla $\H_\A\simeq\C^D$~\cite{Schoen2005} interacting sequentially with a sequence of $d$-dimensional qudits $\H_\B\simeq\C^d$ (each initially in $\ket0$). 
In each step of the protocol, a new qudit is attached to the chain and a unitary is performed on the joint qudit-ancilla system $\H_\A\otimes\H_\B$.
The whole operation can expressed as an isometry $V: \H_\A\to\H_\A\otimes\H_\B$.
If $\{\ket i\}, i\in\{0, \dots, d-1\}$ and $\{\ket{\alpha}\}, \alpha\in\{0, \dots, D-1\}$ are respectively bases for qudit and ancilla, we can write
\begin{equation}
	 V = \sum_{i=0}^{d-1}\sum_{\alpha,\beta=0}^{D-1}V_{\alpha\beta}^i\ket{i}_\P\ket{\alpha}_\V\bra\beta.
	\label{eq:isometry}
\end{equation}
Here and in the rest of the article, we use Latin superscript letters for physical indices/legs, Greek subscript letters for the virtual (ancilla) indices/legs and a subscript $\P$ ($\V$) to label states in the physical (virtual) Hilbert space.
The isometry condition implies that contracting the outgoing legs with the corresponding legs on the conjugate tensor yields the identity on the remaining legs, \emph{viz.}
\begin{equation}
	\sum_{i=0}^{d-1} V^{i\dagger} V^i =
    \begin{array}{c}
		\begin{tikzpicture}[scale=.4,thick,decoration={
		  		  	markings, mark=at position 0.7 with {\arrow{>}}}]
			 \draw (0, 0) -- (0, 2);
			\foreach \x in {0,1,...,1}{
				\begin{scope}[shift={(0, 2*\x)}]
			 	 	\draw[postaction={decorate}, color=ancilla] (-3/2, 0)--(-1/2, 0);
			 	 	 \draw[postaction={decorate}, color=ancilla] (1/2, 0)--(1, 0);
					\filldraw[fill=tensor] (-1/2,-1/2) rectangle (1/2,1/2);
			 	 \end{scope}
			 }
			 \draw[color=ancilla] (1, 0) -- (1, 2);
		\end{tikzpicture}
    \end{array}
    =
    \begin{array}{c}
		\begin{tikzpicture}[scale=.4,thick]
			\draw[color=ancilla] (0, 0) -- (1, 0) -- (1, 2) -- (0, 2);
		\end{tikzpicture}
    \end{array}
    = \1,
	\label{eq:isometry_condition}
\end{equation}
where we also introduced graphical notation in which tensors are represented by boxes and their indices/legs by lines.
Here and in the following, we draw the physical legs in black and the ancilla/virtual legs in red, and we graphically depict the space the isometry acts on (the input) by legs with ingoing arrows, and the target space (the output) by outgoing arrows.
Applying $N$ isometries, and requiring that final state be a product state of ancilla and physical qudits, the reduced state of the qudits is an MPS in canonical form. Conversely, any MPS can be written in this way~\cite{Schoen2005}.

For later convenience, we allow the ancilla dimension to vary across the system (as a function of time).
In particular, we truncate the ancilla dimension such that the reduced state of the ancilla is always full rank, which means that before the first step and after the last step, the ancilla dimension is $1$.
This allows us to write the MPS as
\begin{equation}
	\begin{aligned}
		\ket{\psi_\mathrm{MPS}}&=\sum_{i_1\dots i_N=0}^{d-1} V_{[n]}^{i_n}\cdots V_{[1]}^{i_1}\ket{i_N\dots i_1}\\
		&=
    	\begin{array}{c}
			\begin{tikzpicture}[scale=.35,thick,decoration={
		  		  		markings, mark=at position 0.8 with {\arrow{>}}}]
				\foreach \x in {1,2,...,2}{
					\draw[postaction={decorate},color=ancilla] (2*\x-3/2, -1/2) -- (2*\x-0.5, -1/2);
				}
				\foreach \x in {1,2,...,2}{
					\begin{scope}[shift={(2*\x-2, 0)}]
						\filldraw[fill=tensor] (-1/2,-1) rectangle (1/2,1);
						\draw[postaction={decorate}] (1,1/2)--(1,1.25);
						\draw (1/2,1/2)--(1,1/2);
						\draw (1, 1.8) node {$i_\x$};
					\end{scope}
				}
				\draw[dotted] (4, 0) -- (5, 0);
				\begin{scope}[shift={(7, 0)}]
					\filldraw[fill=tensor] (-1/2,-1)--(-1/2,1)--(1/2,1)--(1/2,-1)--(-1/2,-1);
					\draw[postaction={decorate}] (1,1/2)--(1,1.25);
					\draw (1/2,1/2)--(1,1/2);
					\draw (1, 1.8) node {$i_N$};
				\end{scope}
				\draw[postaction={decorate},color=ancilla] (5.5,-1/2)--(6.5,-1/2);
			\end{tikzpicture}
    	\end{array}.
	\end{aligned}
	\label{eq:MPS}
\end{equation}
Notice that the first isometry (leftmost) is in fact just a state (the input dimension is 1) and the last isometry is a unitary (input and output dimension agree).

During the generation process, the reduced state of the ancilla evolves under the action of the channels obtained by tracing over the chain qudits.
In the $n$th step, the channel reads
\begin{equation}
	\Phi_n: \rho\mapsto
	 \tr_\P( V_{[n]} \rho  V_{[n]}\dagg)
	= \sum_{i=0}^{d-1}V^i_{[n]}\rho V^{i\dagger}_{[n]}.
	\label{eq:channels}
\end{equation}
This allows us to calculate the expectation value of an observable $\O$ on the $n$th site as 
\begin{equation}
	\bra{\psi_\mathrm{MPS}}\O_n\ket{\psi_\mathrm{MPS}} = \tr( V_{[n]}\rho_{n-1} V_{[n]}\dagg\O).
	\label{eq:local_expect}
\end{equation}
Equivalently, we could say that the transfer matrix of the MPS is a CPTP map.

\subsection{Isometric tensor network states}
The isometry property can be generalized to define isometric tensor network states two dimensions~\cite{Zaletel2020}.
We can think of an $N_x\times N_y$-dimensional isoTNS as being produced by $N_x+N_y$ ancillas.
The ancillas fly through each other in the following way: $N_y$ are arranged as a vertical stack and fly left to right, and $N_x$ are arranged as a horizontal stack and fly bottom to top.
Whenever their worldlines cross, we apply an isometry that acts on the two ancillas and generates a physical qudit.
For $(N_x,N_y)=(4,2)$, the process and state can be illustrated as
\begin{equation}
	\begin{aligned}
		\ket{\psi}&=
		\prod^\leftarrow_{m,n}  V_{[mn]}
		=
    	\begin{array}{c}
			\begin{tikzpicture}[scale=.4,thick,decoration={
		  		  		markings, mark=at position 0.8 with {\arrow{>}}}]
				\def\Nx{4}
				\def\Ny{2}
				\tikzmath{\N=int(\Nx-1);\M=int(\Ny-1);}
				\foreach \x in {0,1,...,\N}{
					\foreach \y in {0,1,...,\M}{
						\begin{scope}[shift={(2*\x, 2*\y)}]
							\ifnum \x<\N {
								\draw[postaction={decorate}, color=ancilla] (1/2, 0) -- (3/2, 0);}
							\fi
							\ifnum \y<\M {
								\draw[postaction={decorate}, color=ancilla] (0, 1/2) -- (0, 3/2);}
							\fi
							\filldraw[fill=tensor] (-1/2,-1/2) rectangle (1/2,1/2);
						 	\draw[postaction={decorate}] (0, 0) -- (.6, 1.1);
					 	\end{scope}
					}
				}
			\end{tikzpicture}
    	\end{array}.
	\end{aligned}
    \label{eq:isoTNS}
\end{equation}
Here we have used the symbol $\overleftarrow{\prod}$ to denote that the isometries need to be applied in the order of the arrows, i.e., isometries earlier in the worldline are applied first and appear further right in the product.
Note that in our convention the arrows point in the direction of time and opposite to those in Ref.~\cite{Zaletel2020}.
IsoTNS possess an orthogonality centre, which is the site generated by the first isometry and correspondingly the origin of the flow in the virtual legs, as indicated by the arrows.
In general, the orthogonality centre can lie anywhere in the lattice~\cite{Zaletel2020}, but here we only consider states where it is in the bottom left corner, as in \cref{eq:isoTNS}.

As for MPS, expectation values can be computed by following the time evolution of the ancillas until the required row and column as in \cref{eq:channels}.
The reduced state of the ancillas $\rho_\V$ after tracing the physical degree of freedom evolves under the channels [\emph{cf.} \cref{eq:channels}]
\begin{equation}
	\begin{aligned}
		\Phi_{mn}: \rho_\V &\mapsto \tr_\P( V_{[mn]} \rho_\V  V_{[mn]}\dagg)=\sum_{i=0}^{d-1}V^i_{[mn]}\rho_\V V^{i\dagger}_{[mn]}\\
		&=
    	\begin{array}{c}
			\begin{tikzpicture}[scale=.4,thick,decoration={
		  		  		markings, mark=at position 0.8 with {\arrow{>}}}]
			 	\draw[postaction={decorate}] (0, -1/2) -- (-3/2, -1/2);
				\draw (-3/2,-1/2)--(-3/2,-2)--(4+3/2,-2)--(4+3/2,-1/2);
			 	\draw[postaction={decorate},color=ancilla] (0,.3)--(-3/2,.3);
			 	\draw[postaction={decorate},color=ancilla] (0,.7)--(-3/2,.7);
			 	\begin{scope}[shift={(2,0)}]
			 		\draw[postaction={decorate},color=ancilla] (0,.3)--(-3/2,.3);
			 		\draw[postaction={decorate},color=ancilla] (0,.7)--(-3/2,.7);
			 		\draw[postaction={decorate},color=ancilla] (1/2,.3)--(3/2,.3);
			 		\draw[postaction={decorate},color=ancilla] (1/2,.7)--(3/2,.7);
			 		\draw[postaction={decorate},color=ancilla] (2+1/2,.3)--(2+3/2,.3);
			 		\draw[postaction={decorate},color=ancilla] (2+1/2,.7)--(2+3/2,.7);
			 	\end{scope}
				\filldraw[fill=tensor] (-1/2,-1) rectangle (1/2,1);
				\draw (0, 0) node {\scriptsize $V$};
				\filldraw[color=black, fill=white, thick](2, 1/2) circle (0.7);
				\draw (2, 1/2) node {\scriptsize $\rho_\V$};
				\begin{scope}[shift={(4,0)}]
			 		\draw[postaction={decorate}] (0, -1/2) -- (3/2, -1/2);
					\filldraw[fill=tensor] (-1/2,-1) rectangle (1/2,1);
					\draw (0, 0) node {\scriptsize $V\dagg$};
				\end{scope}
			\end{tikzpicture}
    	\end{array},
	\end{aligned}
	\label{eq:iso_channel}
\end{equation}
where in the illustration we have suppressed indices and the rest of the legs of $\rho$ for clarity.
Here, $V_{[mn]}$ acts on the reduced state of two incoming ancillas (the $n^{\mathrm{th}}$ ancilla from the vertical stack and the $m^{\mathrm{th}}$ ancilla from the horizontal stack).
To evaluate the expectation value of an observable $\O$ at site $(m,n)$ we need to first apply all ``earlier'' channels
\begin{equation}
	\bra\psi\O_{mn}\ket\psi=\tr_\P\!\left[ V_{mn}\left(\prod_{\substack{k,l\leq m,n,\\(k,l)\neq(m,n)}}^{\leftarrow}\!\!\!\!\!\Phi_{kl}(\ket{\vec 0}\bra{\vec 0})\right) V_{mn}\dagg \O\right].
	\label{eq:iso_local_expect}
\end{equation}
Note that this is merely a reinterpretation of tensor network contraction, but where the fact that we have isometries allows us to interpret the contraction as CPTP channels acting on the virtual degrees of freedom.

The definition of isoTNS can straightforwardly be extended to arbitrary directed acyclic graphs and with mixed bond dimensions, but for the rest of this work we focus on the physically relevant two-dimensional states.

\subsection{Projected-entangled pair states and injectivity}
Projected-entangled pair states (PEPS) is an alternative way to think about tensor network states.
\begin{definition}[PEPS]
	Given a graph $G=(W,E)$ with vertices $W$ and edges $E$, we associate to each edge $e\in E$ an (unnormalized) entangled pair state $\ket{\phi_e}=\sum_i^{D_e}\ket{ii}$, where $D_e$ is called \emph{bond dimension}. This maximally entangled pair is distributed to the two vertices connected by the edge.
	As a result, we obtain the state $\ket\Phi=\bigotimes_e\ket{\phi_e}$, which is defined on a lattice given by the vertices, where each vertex $v$ has the local Hilbert space dimension $\dim{\H_v}=\prod_{e\in E_v}D_e$.
	A PEPS is obtained by applying a linear map to each vertex that maps (``projects'') the halves of the entangled pairs at that vertex to a physical space $P:\C_{\prod_e D_e}\to\C_{d_v}$, where $d_v$ is the physical dimension at vertex $v$.
	\label{def:peps}
\end{definition}
An isoTNS written as in \cref{eq:isoTNS} is a PEPS with projectors
\begin{equation}
    \begin{array}{c}
		\begin{tikzpicture}[scale=.4,thick,decoration={
		  		  	markings, mark=at position 0.8 with {\arrow{>}}}]
			\draw[postaction={decorate}] (0, 0) -- (-3/2, 0);
			\begin{scope}[shift={(3/2,0)}]
			 	\draw[postaction={decorate},color=ancilla] (-3/2,-.3)--(0,-.3);
			 	\draw[postaction={decorate},color=ancilla] (-3/2,-.7)--(0,-.7);
			 	\draw[postaction={decorate},color=ancilla] (0,.3)--(-1,.3);
			 	\draw[postaction={decorate},color=ancilla] (0,.7)--(-1,.7);
			\end{scope}
			\filldraw[fill=tensor] (-1/2,-1) rectangle (1/2,1);
			\draw (0, 0) node {\scriptsize $P$};
		\end{tikzpicture}
    \end{array}
	= P^k_{\alpha\beta\gamma\delta} = V^{k}_{\alpha\beta,\gamma\delta}
	=
    \begin{array}{c}
		\begin{tikzpicture}[scale=.4,thick,decoration={
		  		  	markings, mark=at position 0.8 with {\arrow{>}}}]
			\draw[postaction={decorate}] (0, -1/2) -- (-3/2, -1/2);
			\draw[postaction={decorate},color=ancilla] (0,.3)--(-3/2,.3);
			\draw[postaction={decorate},color=ancilla] (0,.7)--(-3/2,.7);
			\begin{scope}[shift={(3/2,0)}]
			 	\draw[postaction={decorate},color=ancilla] (0,.3)--(-1,.3);
			 	\draw[postaction={decorate},color=ancilla] (0,.7)--(-1,.7);
			\end{scope}
			\filldraw[fill=tensor] (-1/2,-1) rectangle (1/2,1);
			\draw (0, 0) node {\scriptsize $V$};
		\end{tikzpicture}
    \end{array},
	\label{eq:P}
\end{equation}
Thus, $P$ and $V$ are the same tensor, but $V$ is thought of as a map from $D^2$ to $d*D^2$, whereas $P$ is a map from $D^4$ to $d$.

The most studied condition that can be imposed on PEPS is \emph{injectivity}.
Injective PEPS are the unique ground states of local frustration-free Hamiltonians~\cite{Perez-Garcia2008}.
They have the property that every tensor has an inverse acting on the physical index.
Thus, we require that $d\geq D^4$ and that every PEPS projector $P_{[ij]}$ has an inverse $P^{-1}_{[ij]}$ such that
\begin{equation}
	\sum_{i=1}^{d}\left( P^{-1}_{[ij]} \right)_{\alpha\beta\gamma\eta}^i(P_{[ij]})_{\alpha'\beta'\gamma'\eta'}^i=
	\delta_{\alpha\alpha'}\delta_{\beta\beta'}\delta_{\gamma\gamma'}\delta_{\eta\eta'}.
	\label{eq:injectivity}
\end{equation}
Graphically, this condition reads
\begin{equation}
    \begin{array}{c}
		\begin{tikzpicture}[scale=.4,thick,decoration={
		  		  	markings, mark=at position 0.8 with {\arrow{>}}}]
			\draw[postaction={decorate}] (0, 0) -- (-3/2, 0);
			\begin{scope}[shift={(3/2,0)}]
			 	\draw[postaction={decorate},color=ancilla] (-3/2,-.3)--(0,-.3);
			 	\draw[postaction={decorate},color=ancilla] (-3/2,-.7)--(0,-.7);
			 	\draw[postaction={decorate},color=ancilla] (0,.3)--(-1,.3);
			 	\draw[postaction={decorate},color=ancilla] (0,.7)--(-1,.7);
			\end{scope}
			\filldraw[fill=tensor] (-1/2,-1) rectangle (1/2,1);
			\draw (0, 0) node {\scriptsize $P$};
			\begin{scope}[xscale=-1,shift={(3,0)}]
				\draw[postaction={decorate}] (0, 0) -- (-3/2, 0);
				\begin{scope}[shift={(3/2,0)}]
			 		\draw[postaction={decorate},color=ancilla] (-3/2,-.3)--(0,-.3);
			 		\draw[postaction={decorate},color=ancilla] (-3/2,-.7)--(0,-.7);
			 		\draw[postaction={decorate},color=ancilla] (0,.3)--(-1,.3);
			 		\draw[postaction={decorate},color=ancilla] (0,.7)--(-1,.7);
				\end{scope}
				\filldraw[fill=tensor] (-1/2,-1) rectangle (1/2,1);
			\end{scope}
			\draw (-3, 0) node {\scriptsize $P^{\text-1}$};
		\end{tikzpicture}
    \end{array}
    =
    \begin{array}{c}
		\begin{tikzpicture}[scale=.4,thick]
			\draw[color=ancilla] (3,-.3)--(0,-.3);
			\draw[color=ancilla] (3,-.7)--(0,-.7);
			\draw[color=ancilla] (3,.3)--(0,.3);
			\draw[color=ancilla] (3,.7)--(0,.7);
		\end{tikzpicture}
    \end{array}.
	\label{eq:injectivity_graphical}
\end{equation}
To avoid pathological situations in which tensors are arbitrarily close to non-injective ones, we demand that all singular values of $P$ are greater or equal to some $\delta>0$, or equivalently that
$\norm{P^{-1}}_\infty\leq 1/\delta$.
A similar condition has been used before \cite{Schwarz2012,Schwarz2017,Anshu2024} and the corresponding states (without the isometry condition) have been dubbed $\delta$-injective PEPS~\cite{Anshu2024}. (We note that defining the PEPS tensor as acting on \emph{normalized} entangled pair states corresponds to dividing each element of $P$ by a factor $D$ and a corresponding change in $\delta$.)

\subsection{Injective isoTNS and depolarizing noise}
\label{sec:injective_isoTNS}
Injectivity imposed on isoTNS has the interesting consequence that the ancillas are guaranteed to experience some amount of local depolarizing noise.
This is the essential ingredient we need later to prove that sufficiently injective isoTNS are easy to contract.
In the following computations we take the bond dimensions to be $D$ across all bonds for simplicity, but a generalization to inhomogeneous bond dimension is straightforward.

Let us first characterize what values $\delta$ can take.
Since the Frobenius norm is equal to the element-wise 2-norm, we have $\norm{V}_F=\norm{P}_F=\sqrt{\sum_i^{D^4} \sigma_i^2}$, 
where $\{\sigma_i\}$ are the singular values of $P$.
Since $V$ is an isometry mapping a $D^2$-dimensional space to $dD^2$, we have $\norm{V}_F=D$.
Now consider the spectral norm $\norm P_\infty$.
For a fixed Frobenius norm, it is minimized when all its $D^4$ singular values are equal, which in our case gives $1/D\leq\norm P_\infty$.
In turn it is maximized if one singular value is large and all the others are as small as possible, i.e., equal to $\delta$.
Together with the lower bound, this yields
\begin{equation}
	1/D\leq \norm P_\infty\leq \sqrt{D^2-\delta^2(D^4-1)}.
	\label{eq:spectral_bounds}
\end{equation}
Both inequalities can only be satisfied if
\begin{equation}
	\delta\leq1/D.
	\label{delta_condition}
\end{equation}
If this bound is saturated, all singular values of $P$ are $1/D$, which means it is a (rescaled) isometry as well.
(The scaling factor $D$ stems from our definition of PEPS in terms of unnormalized entangled pairs, see \cref{def:peps}.)
States obtained by acting with an isometry on maximally entangled pairs have been considered before and are called \emph{isometric PEPS}~\cite{Schuch2010}.
These PEPS have strictly finite nearest neighbour correlations and can be prepared using a constant-depth circuit.

Away from this limit, for smaller $\delta$, we find that injectivity causes the ancillas to experience depolarizing noise (a similar result was very recently obtained for a special family of PEPS, namely ground states that encode circuits through teleportation~\cite{Anshu2024}).
To establish this, we insert a resolution of the identity inside the trace in \cref{eq:iso_channel}
\begin{equation}
	\1_{d} = (\1_d - \eta M) + \eta M,\qquad
	M = P^{-1\dagger}P^{-1}/D^2.
	\label{eq:resolution_of_the_identity}
\end{equation}
Recall that $P$ maps virtual to physical space, so $M$ maps physical space to physical space.
This allows us to decompose the channel $\Phi_{mn}$ [\cref{eq:iso_channel}] as a sum of a depolarizing channel $\E_\mathrm{depol}$ of strength $\eta$ and another channel $\E_1$
\begin{equation}
	\begin{aligned}
		\Phi_{mn}(\rho) 
		&= \tr_\P[(\1-\eta M) V_{[mn]}\rho V_{[mn]}\dagg] + \eta\1/D^2\\
		& = (1-\eta)\E_1(\rho) + \eta \E_\mathrm{depol}(\rho).
	\end{aligned}
	\label{eq:splitting_channels}
\end{equation}
To see that $\E_1$ is a proper channel note that it can be written in terms of Kraus operators by diagonalizing $M$.
Taking $U$ to be the unitary that diagonalizes $\1-\eta M=U\dagg\Lambda U$ for some diagonal matrix $\Lambda=\mathrm{diag}(\lambda_1,\lambda_2,\dots)$, we have
\begin{equation}
    \E_1(\rho) = (1-\eta)^{-1} \sum_iK^i\rho K^{i\dagger},
\end{equation}
where we define the Kraus operators $K^i = \sqrt{\lambda_i}\bra{i}U V$
and require $\1-\eta M\geq0$ such that $\lambda_i\geq0$.
To see that $\E_1$ is trace preserving, we can check
\begin{equation}
    \sum_iK^{i\dagger} K^i = V\dagg (\1-\eta M) V =
    \1-\frac{\eta}{D^2} V\dagg (P^{-1})\dagg P^{-1} V.
    \label{eq:trace_preserving}
\end{equation}
We can evaluate the second term in \cref{eq:trace_preserving} as
\begin{equation}
    \begin{array}{c}
		\begin{tikzpicture}[scale=.5,thick,decoration={
			markings, mark=at position 0.8 with {\arrow{>}}}]
		\draw[postaction={decorate}] (0, -1/2) -- (-3/2, -1/2);
		\draw[postaction={decorate},color=ancilla] (0,.3)--(-3/2,.3);
		\draw[postaction={decorate},color=ancilla] (0,.7)--(-3/2,.7);
		\begin{scope}[shift={(3/2,0)}]
			\draw[postaction={decorate},color=ancilla] (0,.3)--(-1,.3);
			\draw[postaction={decorate},color=ancilla] (0,.7)--(-1,.7);
		\end{scope}
		\filldraw[fill=tensor] (-1/2,-1) rectangle (1/2,1);
		\draw (0, 0) node {\scriptsize $V$};
		\begin{scope}[xscale=-1,shift={(2,-1.1)}]
			\begin{scope}[shift={(3/2,0)}]
			 	\draw[postaction={decorate},color=ancilla] (-1,-.3)--(-1/2,-.3);
			 	\draw[postaction={decorate},color=ancilla] (-1,-.7)--(-1/2,-.7);
			 	\draw[postaction={decorate},color=ancilla] (-1/2,.3)--(-1,.3);
			 	\draw[postaction={decorate},color=ancilla] (-1/2,.7)--(-1,.7);
			\end{scope}
			\filldraw[fill=tensor] (-1/2,-1) rectangle (1/2,1);
			\draw (0, 0) node {\scriptsize $P^{\text-1}$};
		\end{scope}
		\draw[color=ancilla] (-4.5,.3)--(-1.5,.3);
		\draw[color=ancilla] (-4.5,.7)--(-1.5,.7);

		\begin{scope}[xscale=-1,shift={(6,0)}]
			\draw[postaction={decorate}] (0, -1/2) -- (-3/2, -1/2);
			\draw[postaction={decorate},color=ancilla] (0,.3)--(-3/2,.3);
			\draw[postaction={decorate},color=ancilla] (0,.7)--(-3/2,.7);
			\begin{scope}[shift={(3/2,0)}]
				\draw[postaction={decorate},color=ancilla] (0,.3)--(-1,.3);
				\draw[postaction={decorate},color=ancilla] (0,.7)--(-1,.7);
			\end{scope}
			\filldraw[fill=tensor] (-1/2,-1) rectangle (1/2,1);
			\draw (0, 0) node {\scriptsize $V\dagg$};
			\begin{scope}[xscale=-1,shift={(2,-1.1)}]
				\begin{scope}[shift={(3/2,0)}]
			 		\draw[postaction={decorate},color=ancilla] (-1,-.3)--(-1/2,-.3);
			 		\draw[postaction={decorate},color=ancilla] (-1,-.7)--(-1/2,-.7);
			 		\draw[postaction={decorate},color=ancilla] (-1/2,.3)--(-1,.3);
			 		\draw[postaction={decorate},color=ancilla] (-1/2,.7)--(-1,.7);
				\end{scope}
				\filldraw[fill=tensor] (-1/2,-1) rectangle (1/2,1);
				\draw (0, 0) node {\scriptsize $P^{\text-1\dag}$};
			\end{scope}
		\end{scope}
		\end{tikzpicture}
    \end{array}
    =
    \begin{array}{c}
		\begin{tikzpicture}[scale=.5,thick]
			\draw[color=ancilla] (-2,.7)--(-1,.7)--(-1,-1)--(1,-1)--(1,.7)--(2,.7);
			\draw[color=ancilla] (-.7,.7) rectangle (.7,-.7);
			\begin{scope}[scale=.5]
				\draw[color=ancilla] (-.7,.7) rectangle (.7,-.7);
			\end{scope}
			\begin{scope}[shift={(0,-.4)}]
				\draw[color=ancilla] (-2,.7)--(-1.4,.7)--(-1.4,-1)--(1.4,-1)--(1.4,.7)--(2,.7);
			\end{scope}
    	\end{tikzpicture}
    \end{array}
	\label{eq:TP}
\end{equation}
Thus, $\sum_iK^{i\dagger} K^i=(1-\eta)\1$.
By assumption, $\norm{D^2M}_\infty\leq 1/\delta^2$, so the maximum $\eta$ we can choose is $D^2\delta^2$.

\section{Complexity of computing local expectation values}
\subsection{Worst-case hardness of isoTNS}\label{sec:worst_case_hardness}
To show that computing local expectation values in isoTNS with constant bond dimension is \BQP-complete, first note that following \cref{eq:isoTNS}, isoTNS can efficiently be prepared on quantum computers, which can be used to sample from the state.
For the converse statement, that any problem in \BQP\ can be mapped to isoTNS, we show that one can embed a brickwork circuit into isoTNS. 
For simplicity, we consider $D=2$, $d=4$ and three types of tensor:
(i) a \emph{gate tensor} $G$ that implements a unitary gate on the ancilla degrees of freedom (with arbitrary $ U_\V$ acting only on the ancillas),
\begin{equation}
	 G = \ket{00}_\P U_\V 
	=
    \begin{array}{c}
		\begin{tikzpicture}[scale=.4,thick,rotate=90,yscale=-1,decoration={markings, mark=at position 0.8 with {\arrow{>}}}]
			\draw (-.5, -0.5) node {\scriptsize $\ket0^{\otimes2}$};
			\draw[postaction={decorate}] (0, 0)--(1, 0);
			\draw[postaction={decorate}] (0, -0.5)--(1, -0.5);
			\draw[color=ancilla] (-1, 1.0)--(1, 1.0);
			\draw[color=ancilla] (-1, 1.5)--(1, 1.5);
			\filldraw[fill=tensor] (-1/2,.7) rectangle (1/2,1.8);
			\draw (0, 1.25) node {\scriptsize $U$};
		\end{tikzpicture}
    \end{array}
    =
    \begin{array}{c}
		\begin{tikzpicture}[scale=.4,thick,rotate=45,decoration={markings, mark=at position 0.8 with {\arrow{>}}}]
			\draw[color=white] (-1/2,-1)--(-1/2,2);
			\draw[postaction={decorate}, color=ancilla] (1/2, 0) -- (3/2, 0);
			\draw[postaction={decorate}, color=ancilla] (0, 1/2) -- (0, 3/2);
			\draw[postaction={decorate}, color=ancilla] (-3/2, 0) -- (-1/2, 0);
			\draw[postaction={decorate}, color=ancilla] (0,-3/2) -- (0,-1/2);
			\filldraw[fill=tensor] (-1/2,-1/2) rectangle (1/2,1/2);
			\draw[postaction={decorate}] (.3,.3) -- (.9, .9);
			\draw (0,0) node {\scriptsize $G$};
		\end{tikzpicture}
    \end{array}
	\label{eq:gate_tensor}
\end{equation}
(ii) a \emph{swap tensor} that swaps the ancillas into the physical space
\begin{equation}
	 S = \delta^i_\alpha\delta^j_\beta\ket{ij}_\P\,{}_\V\bra{\alpha\beta}
	=
    \begin{array}{c}
		\begin{tikzpicture}[scale=.35,thick,decoration={markings, mark=at position 0.8 with {\arrow{>}}}]
			\draw[color=ancilla, postaction={decorate}] (0, -1)--(0,0.3);
			\draw[color=ancilla, postaction={decorate}] (0.5, -1)--(0.5,0.3);
			\draw (0, 1)--(0,0);
			\draw (0.5, 1)--(0.5,0);
		\end{tikzpicture}
    \end{array}
    =
    \begin{array}{c}
		\begin{tikzpicture}[scale=.4,thick,rotate=45,decoration={markings, mark=at position 0.8 with {\arrow{>}}}]
			\draw[color=white] (-1/2,-1)--(-1/2,2);
			\draw[postaction={decorate}, color=ancilla] (1/2, 0) -- (3/2, 0);
			\draw[postaction={decorate}, color=ancilla] (0, 1/2) -- (0, 3/2);
			\draw[postaction={decorate}, color=ancilla] (-3/2, 0) -- (-1/2, 0);
			\draw[postaction={decorate}, color=ancilla] (0,-3/2) -- (0,-1/2);
			\filldraw[fill=swap] (-1/2,-1/2) rectangle (1/2,1/2);
			\draw[postaction={decorate}] (.3,.3) -- (.9, .9);
			\draw (0,0) node {\scriptsize $S$};
		\end{tikzpicture}
    \end{array}
	\label{eq:swap_tensor}
\end{equation}
(note that the red outgoing legs in the last illustration have dimension 1 and are drawn just for visual consistency),
(iii) and an \emph{identity tensor} that simply swaps the ancillas and initializes the physical qubits in $\ket0$
\begin{equation}
	 I = \ket{00}_\P\mathrm{SWAP}_\V
	=
    \begin{array}{c}
		\begin{tikzpicture}[scale=.35,rotate=90,yscale=-1,thick,decoration={markings, mark=at position 0.8 with {\arrow{>}}}]
			\draw[color=white] (-1/2,-1)--(-1/2,2);
			\draw[postaction={decorate}] (0, 0)--(1, 0);
			\draw[postaction={decorate}] (0,-0.5)--(1,-0.5);
			\draw[color=ancilla,postaction={decorate}] (-1, 1.0)--(-.2,1)--(.2,1.5)--(1, 1.5);
			\draw[color=ancilla,postaction={decorate}] (-1, 1.5)--(-.2,1.5)--(.2,1)--(1, 1);
			\draw (-.5,-.5) node {\scriptsize $\ket0^{\otimes2}$};
		\end{tikzpicture}
    \end{array}
    =
    \begin{array}{c}
		\begin{tikzpicture}[scale=.4,thick,rotate=45,decoration={markings, mark=at position 0.8 with {\arrow{>}}}]
			\draw[color=white] (-1/2,-1)--(-1/2,2);
			\draw[postaction={decorate}, color=ancilla] (1/2, 0) -- (3/2, 0);
			\draw[postaction={decorate}, color=ancilla] (0, 1/2) -- (0, 3/2);
			\draw[postaction={decorate}, color=ancilla] (-3/2, 0) -- (-1/2, 0);
			\draw[postaction={decorate}, color=ancilla] (0,-3/2) -- (0,-1/2);
			\filldraw[fill=white] (-1/2,-1/2) rectangle (1/2,1/2);
			\draw[color=ancilla, thin] (0,1/2)--(-1/2,0);
			\draw[color=ancilla, thin] (1/2,0)--(0,-1/2);
			\draw[postaction={decorate}] (.3,.3) -- (.9, .9);
			\draw (0,0) node {\scriptsize $I$};
		\end{tikzpicture}
    \end{array}.
	\label{eq:identity_tensor}
\end{equation}
Given any one-dimensional brickwork circuit with nearest-neighbour gates, we can now define a corresponding isoTNS using the following procedure:
(i) replace all unitaries by the corresponding gate tensor $G$, (ii) at the end of the circuit place a layer of swap tensors $S$, (iii) fill up the remaining sites with the identity tensor to obtain a square lattice geometry.
For example for a circuit of four qubits and depth five, we have (suppressing physical legs for clarity)
\begin{equation}
	\begin{aligned}
		\ket{\psi}&=
    	\begin{array}{c}
			\begin{tikzpicture}[scale=.3,thick,rotate=45,decoration={
		  		  		markings, mark=at position 0.8 with {\arrow{>}}}]
				\def\N{4}
				\def\M{4}
				\foreach \x in {0,1,...,\N}{
					\foreach \y in {0,1,...,\M}{
						\begin{scope}[shift={(2*\x, 2*\y)}]
							\ifnum \x<\N {
								\draw[postaction={decorate}, color=ancilla] (1/2, 0) -- (3/2, 0);}
							\fi
							\ifnum \y<\M {
								\draw[postaction={decorate}, color=ancilla] (0, 1/2) -- (0, 3/2);}
							\fi
							\filldraw[fill=white] (-1/2,-1/2) rectangle (1/2,1/2);
							\draw (0,0) node {\scriptsize $I$};
							\ifnum \numexpr\x+\y>0
							\ifnum \numexpr\x+\y<\numexpr\N+2
								\ifnum \numexpr\x-\y<2
								\ifnum \numexpr\y-\x<2
									{\filldraw[fill=tensor] (-1/2,-1/2) rectangle (1/2,1/2);
									\draw (0,0) node {\scriptsize $G$};}
							\fi\fi\fi\fi
							\ifnum\numexpr\x+\y=\numexpr\N+2
								{\filldraw[fill=swap] (-1/2,-1/2) rectangle (1/2,1/2);
								\draw (0,0) node {\scriptsize $S$};}
							\fi
						\end{scope}
					}
				}
			\end{tikzpicture}
    	\end{array}.
	\end{aligned}
    \label{eq:BQP-circuit}
\end{equation}
The size of the isoTNS is linear in both number of qubits and depth and the reduced state of the physical qudits corresponding to sites with the swap tensor (marked in yellow) is the output of the circuit.
Computing local expectation values on these sites is therefore \BQP-complete.

\subsection{Hardness of isoTNS with a translationally invariant bulk tensor}

Here we consider isoTNS where all tensors are the same isometry from two ancillas to two ancillas and the physical qudit
\begin{equation}
	V_0:\H_A^{\otimes 2}\to\H_A^{\otimes 2}\otimes\H_P
	\label{eq:V-TI}
\end{equation}
and show that computing expectation values can still be hard (dependent on boundary conditions).

Before we get there, we first need to specify the boundary conditions of the state, which turn out to be very important.
Placing the isometry \cref{eq:V-TI} on every lattice site, we get a tensor network with dangling virtual indices around the perimeter, which is an isometry $W_{N,M}$ from $N+M$ input legs to $N+M$ output legs and $NM$ physical legs, $W:\H_A^{\otimes(N+M)}\to\H_A^{\otimes(N+M)}\otimes\H_P^{\otimes NM}$, which can be drawn as
\begin{equation}
	\begin{aligned}
		W&=
    	\begin{array}{c}
			\begin{tikzpicture}[scale=.4,thick,rotate=45,decoration={
		  		  		markings, mark=at position 0.8 with {\arrow{>}}}]
				\def\N{2}
				\def\M{2}
				\foreach \x in {0,1,...,\N}{
					\foreach \y in {0,1,...,\M}{
						\begin{scope}[shift={(2*\x, 2*\y)}]
							\draw[postaction={decorate}, color=ancilla] (1/2, 0) -- (3/2, 0);
							\draw[postaction={decorate}, color=ancilla] (0, 1/2) -- (0, 3/2);
							\draw[postaction={decorate}, color=ancilla] (-3/2, 0) -- (-1/2, 0);
							\draw[postaction={decorate}, color=ancilla] (0, -3/2) -- (0, -1/2);
							\filldraw[fill=white] (-1/2,-1/2) rectangle (1/2,1/2);
							\draw[postaction={decorate}] (0.2, 0.2) -- (.8, .8);
							\draw (0,0) node {\scriptsize $V_0$};
						\end{scope}
					}
				}
			\end{tikzpicture}
    	\end{array}.
	\end{aligned}
    \label{eq:TI-isoTNS}
\end{equation}
Imposing periodic boundary conditions would correspond to connecting legs on opposite sides.
However, this removes essentially all favourable properties of isoTNS, including the notion of an orthogonality centre, their sequential preparability and the fact that their norm is 1.
For this reason, also the original definition of isoTNS only considers open boundary conditions~\cite{Zaletel2020}.

Thus, we are forced to consider different choices of open boundary conditions.
We have to additionally distinguish between dangling legs that are \emph{inputs} and those that are \emph{outputs} of isometries.
We consider a boundary condition where the input legs are contracted with a product state $\ket{\pi_0}$ and the output legs are promoted to physical legs. This situation corresponds to a standard isoTNS that can be thought of as being generated in a physical process. The state obtained can be illustrated as (here for $(N_x,N_y)=(4,2)$)
\begin{equation}
	\begin{aligned}
		\ket{\psi^{(i)}}_{P\cup A}&=
		W\ket{\pi_0}_A=
    	\begin{array}{c}
			\begin{tikzpicture}[scale=.4,thick,decoration={
		  		  		markings, mark=at position 0.8 with {\arrow{>}}}]
				\def\Nx{4}
				\def\Ny{2}
				\tikzmath{\N=int(\Nx-1);\M=int(\Ny-1);}
				\foreach \x in {0,1,...,\N}{
					\foreach \y in {0,1,...,\M}{
						\begin{scope}[shift={(2*\x, 2*\y)}]
							\draw[postaction={decorate}, color=ancilla] (1/2, 0) -- (3/2, 0);
							\draw[postaction={decorate}, color=ancilla] (0, 1/2) -- (0, 3/2);
							\filldraw[fill=tensor] (-1/2,-1/2) rectangle (1/2,1/2);
						 	\draw[postaction={decorate}] (0, 0) -- (.6, 1.1);
							\ifnum \x=\N {
								\draw[postaction={decorate}] (3/2,0) -- (3/2+.3, .55);}
							\fi
							\ifnum \y=\M {
								\draw[postaction={decorate}] (0,3/2) -- (.3, 3/2+.55);}
							\fi
					 	\end{scope}
					}
				}
			\end{tikzpicture}
    	\end{array}.
	\end{aligned}
    \label{eq:case-i}
\end{equation}
For the family of problems where the input state at the boundary, $\ket{\pi_0}$, can be a translationally varying product state, computing local expectation values in such a state remains \BQP-hard in the worst case, which we establish by embedding arbitrary quantum computation, as before. 

To establish hardness, we interpret the isoTNS as a translation-invariant 1D quantum circuit. It has previously been recognized that, as long as the input to such a state can be a translationally varying product state, the circuit can be translation-invariant by encoding a Turing machine into the circuit~\cite{Gottesman2009}. Here, we have to adapt this result to the present case, where the subtlety is that for a square isoTNS, the circuit has an oblique shape. We recap the Turing machine embedding into a translation-invariant circuit and detail its encoding into the iso TNS in \cref{app:ti-isotns}.

It will be interesting to investigate whether this also implies that algorithms to contract infinite tensor network states that are based on approximating the transfer matrix must fail, but we leave this connection to future work.

\subsection{Strongly injective isoTNS can be contracted efficiently}
\label{sec:complexity_transition}
\Cref{eq:BQP-circuit} defines a highly artificial state and one might wonder what happens to states that realistically appear in the study of ground states.
In the absence of topological order, PEPS are typically injective (or normal), which is equivalent to the state being the unique ground state of a nearest-neighbour frustration-free Hamiltonian~\cite{Perez-Garcia2008}.
Intuitively, injectivity means that every observable on the virtual space can be accessed using a suitably defined observable on the physical space and as we have shown, this leads to depolarizing noise.  
Depolarizing noise reduces the computational power of the ancillas, eventually leading to a transition to a classically simulable regime, of a type first predicted more than 20 years ago~\cite{Aharonov1996,Aharonov2000}.
We use this to show two things: (i) strongly injective isoTNS can be contracted efficiently, and (ii) weakly injective isoTNS are still \BQP-complete.

To find a provably efficient classical algorithm to compute a local expectation value $\langle \O \rangle$, we adapt a method due to Aharonov~\cite{Aharonov2000} to map contracting the tensor network to percolation.
The key insight is that using \cref{eq:splitting_channels}, we can think of the contraction in $\langle \O \rangle$ as a partition sum over all possible assignments of $\E_1=(\Phi-\eta\E_\mathrm{depol})/(1-\eta)$ or $\E_\mathrm{depol}$ to the sites of the isoTNS.
Rather than computing the full partition sum, we use Monte Carlo sampling to obtain an estimate with a typical error that decays in the number of samples $M$ as $1/\sqrt{M}$.
To make the algorithm efficient, we only accept a sample if the cluster size lies below some cutoff $s_\mathrm{th}$.
\begin{minipage}{.94\linewidth}
\begin{algorithm}[H]
	\caption{Contraction algorithm.}\label{alg}
	\begin{algorithmic}[1]
		\Require Observable $\O$ with strictly local support $\supp(\O)$, isoTNS with injectivity $\eta$, cutoff $s_\mathrm{th}$, number of samples $M$.
		\For{$i\in [M]$}
			\State Assign a channel to each site outside $\supp(\O)$: assign $\E_{\mathrm{depol}}$ with probability $\eta$ (occupied) and otherwise $\E_1$.
			\State Evaluate the size $\mathcal S$ of the cluster connected to the support of the observable.
			\If{$\mathcal S>s_{\mathrm{th}}$}
				\State \texttt{Continue.}
			\Else
				\State Contract the cluster exactly, store result $x_i$.
			\EndIf
		\EndFor
		\State Compute average $\bar x$ of all $x_i$.
		\State \Return $\bar x$
	\end{algorithmic}
\end{algorithm}
\end{minipage}

We illustrate a typical sample of the above algorithm in \cref{fig:percolation}. Note that to compute the observable in a given sample, we only need to contract the connected cluster of filled sites (see \cref{fig:percolation}).
If $\eta\geq 0.41$, then the equivalent site percolation problem is subcritical (i.e.~the occupied/filled sites do not percolate) and the typical size of the cluster connected to the site of the local observable is independent of the system size~\cite{StaufferAharony1992}.
More precisely, large clusters are exponentially suppressed in size~\cite{StaufferAharony1992}:
on a lattice of $n$ sites and when the effective 2D percolation problem is subcritical, the size of the cluster connected to the local observable, $\mathcal{S}$, as a random variable satisfies
\begin{equation}
	\text{Prob}(\mathcal{S} \geq s) \leq O(s^2\exp(-s/\xi_0)),
\end{equation}
for some constant $\xi_0$~\cite{Bazant2000}.

By discarding all samples in which the connected cluster is larger than $s_\mathrm{th}$, we introduce a systematic error.
This error is upper bounded by the fraction of discarded samples, which we can bound as
\begin{align}
&\left|{\mathbb{E}(\langle \O \rangle | \mathcal{S} \leq s_\text{th}) - \langle \O \rangle}\right| \nonumber\\
&\qquad=\frac{\left| \langle \O\rangle - \mathbb{E}(\langle \O \rangle|\mathcal{S} \geq s_\text{th}) \text{Prob}(\mathcal{S} \geq s_\text{th})\right|}{1 - \text{Prob}(\mathcal{S} \geq s_\text{th})} \nonumber\\
&\qquad \leq 2\norm{\O} O(s_\text{th}^2 \exp(-s_\text{th}/\xi_0)),
\end{align}
where we have used the fact that $\left|\langle \O \rangle\right |, \left |\mathbb{E}(\langle \O \rangle | \mathcal{S}\geq s_\text{th}) \right | \leq \norm{\O}$. Thus, to compute $\langle \O \rangle$ to a precision $\eps$, choosing a cluster size threshold $s_\text{th} = \Theta(\log(\eps^{-1}))$ is sufficient.
This yields a run-time of $\text{poly}(\eps^{-1})$ for contracting each individual accepted sample, and $\Theta(\eps^{-2})$ samples are needed to reduce the variance of the estimated average to $O(\eps)$.
Putting these two estimates together, we conclude that, in the subcritical regime, \cref{alg} has a run time independent of system size and inverse-polynomial in the desired precision and is thus efficient.

\begin{figure}[t!]
	\begin{tikzpicture}[scale=.3,thick,rotate=45,decoration={
		  		markings, mark=at position 0.7 with {\arrow{>}}}]
		\def\N{4}
		\def\M{4}
		\def\a{.5}
		\foreach \x/\y in {2/2,1/2,2/1} {
			\begin{scope}[shift={(2*\x, 2*\y)}]
				\fill[lightgray](-2*\a,-2*\a) rectangle (2*\a,2*\a);
			\end{scope}
		}
		\foreach \x in {0,1,...,\N}{
			\foreach \y in {0,1,...,\M}{
				\begin{scope}[shift={(2*\x, 2*\y)}]
					\ifnum \x<\N {\draw[postaction={decorate}, color=ancilla] (1/2, 0) -- (3/2, 0);}\fi
					\ifnum \y<\M {\draw[postaction={decorate}, color=ancilla] (0, 1/2) -- (0, 3/2);}\fi
				\end{scope}
			}
		}
		\foreach \x/\y in {4/4,4/3,4/0,3/4,3/1,3/0,2/3,2/2,2/1,1/3,1/2,0/1} {
			\begin{scope}[shift={(2*\x, 2*\y)}]
				\filldraw[fill=myblue](-\a,-\a) rectangle (\a,\a);
			\end{scope}
		}
		\begin{scope}[shift={(2*2, 2*2)}]
			\filldraw[fill=ancilla](-\a,-\a) rectangle (\a,\a);
		\end{scope}
	\end{tikzpicture}
	\caption{Illustration of one sample in the algorithm to compute local expectation values.
		Each site is randomly assigned the depolarizing channel (empty crossing) or the channel $\E_1$ (blue tensor).
		In the easy phase, sites with tensors do not percolate and clusters are typically small.
		As a result, to compute an observable on the red site, we only need to contract the cluster connected to the sites (blue) and in fact only the part of the cluster in the past light cone (grey background).
	}
	\label{fig:percolation}
\end{figure}

\subsection{Weakly injective isoTNS are still hard}
\label{sec:hard_regime}

Next, we consider the question of hardness of computing local observables in injective isoTNS for any $\delta$.
Building on the idea in Ref.~\cite{Anshu2024} and from the fact that a 1D quantum circuit with nearest-neighbour gates can be fault tolerant~\cite{Aharonov1999,Gottesman2000}, we show that for $\delta$ below some threshold, computing local observables is \BQP-hard.
To encode a 1D fault tolerant quantum circuit into the ancilla dynamics, we need to construct injective and isometric tensors that approximately implement (a) a given two-qubit unitary $U$ and (b) a restart operation on one of the ancilla qubits.

As in \cref{sec:worst_case_hardness}, we take bond dimension $D = 2$, but now we need physical dimension $d = D^4 = 16$ to obtain injective tensors.
First, we perturb the gate tensor \cref{eq:gate_tensor} to injectivity.
To this end, consider a slightly depolarized unitary
\begin{equation}
	\E_U(\rho) = (1 - p) U\rho U^\dagger + p \frac{\1\otimes\1}{4} \tr(\rho).
\end{equation}
As a channel on $\mathbb{C}^D \otimes \mathbb{C}^D$, $\E_U$ has full Kraus rank and thus is described by $(D^2)^2 = D^4$ linearly independent Kraus operators $A^1, A^2 \dots A^{D^4}$.
We can then choose the isometry from the input ancillas to the output ancillas and physical qudit to be the Stinespring dilation of $A^i$
\begin{equation}
	V := \sum_{i = 1}^{D^4} A^i \otimes \ket{i},
\end{equation}
where the states $\ket{1}, \ket{2} \dots \ket{D^4}$ are a basis for the Hilbert space $\mathbb{C}^d$ corresponding to the physical qudit.
Furthermore, as calculated in \cref{app:cond_num_ftc}, if we define the PEPS tensor $P$ in terms of $V$ [see \cref{eq:P}], we have $\norm{P^{-1}}_\infty=1/\delta=2/\sqrt{p}$.

Next, we perturb the restart gate to injectivity.
In the ideal case, it should reset (one of) the ancilla qubits, i.e., implement the map $\rho \to \tr_1(\rho) \otimes \ket{0}\!\bra{0}$.
To make it injective, we again implement a slightly depolarized version of this channel
\begin{equation}
	\E_\text{res}(\rho) = (1 - p) \ket{0}\!\bra{0}\otimes \text{Tr}_1(\rho) + p\frac{\1^{\otimes 2}}{4} \text{Tr}(\rho).
\end{equation}
As with the channel applying the unitary gate, this channel also has full Kraus rank and be described by $D^4$ linearly independent Kraus operators, which can then be mapped to an isometry via the Stinespring dilation, which gives again a PEPS tensor with $\norm{P^{-1}}_\infty=1/\delta=2/\sqrt{p}$.

Now, we use the circuit to isoTNS embedding described in section \ref{sec:worst_case_hardness} and depicted schematically in \cref{eq:BQP-circuit}, but instead of embedding a 1D quantum circuit into it directly, we embed its fault tolerant version (which would include both two-qubit unitaries and restart operations) together with the decoding circuit.
If $\delta$ is small enough, then the noise strength is below the fault tolerance threshold.
We point out that while it may appear as if the identity tensors in the circuit \cref{eq:BQP-circuit} introduce an error, in fact they are only relevant at the boundary of the circuit, when they propagate an ancilla that is part of the circuit.
Thus, they contribute depolarizing noise at the same rate as all the other gates and are corrected automatically.
We conclude that this embedding establishes that computing local observables would again become \BQP-complete when $p$ is small enough, and thus computing local observables on (weakly) injective isoTNS is as hard as on general isoTNS in the worst case.

\section{Complexity of sampling from isoTNS}
\textsc{sampling} is a computational task where we are required to return $K$ samples that are distributed according to the probability distribution $p(\vec x) = |\langle\vec x\ket{\psi}|^2$, where $\ket{\vec x}$ are computational basis states.

Sampling from output states of random circuits with or without depolarizing noise has been studied intensely~\cite{Hangleiter2023} as it was perceived to be a potential avenue to demonstrate quantum computational advantage~\cite{Bouland2018}.
In \cref{sec:hard_sampling} we show that sampling to multiplicative error is hard even from maximally injective isoTNS, which are equal to isometric PEPS.
This result closely mirrors earlier results that establish hardness when sampling from low-depth circuits or cluster states~\cite{Terhal2002,Gao2017}.

Another sampling problem that has been studied intensely in recent years is if instead of depolarizing noise, random unitary circuits are interspersed with measurements.
In this setting, high measurement rates drive the system to an area-law state, whereas for low measurement rates, the system tends to a volume-law state, in a phenomenon that has been dubbed ``measurement-induced phase transition'' (MIPT)~\cite{Skinner2019,Li2019b,Chan2019}.
In the thermodynamic limit, there is a sharp transition between these regimes and this setting has elicited tremendous interest~\cite{Fisher2023}.

At face value, sampling from states and sampling from the measurements added to a dynamical circuits are not directly related, as the former probes properties of the state, and the latter of the dynamics.
However, in line with the central theme of this paper, we can use the fact that isoTNS are closely related to physical dynamics to establish a map between sampling from the isoTNS and monitored quantum dynamics of the ancillas. 
In particular, this allows us to prove the existence of an MIPT when sampling from certain isoTNS as a function of the injectivity parameter.
To show this, we construct a one-parameter family of isoTNS that interpolates from the provably hard regime to a provably easy regime in \cref{sec:MIPT}.

\subsection{Hardness to sample to multiplicative precision}
\label{sec:hard_sampling}
We first consider the problem of sampling an injective isoTNS to multiplicative precision.
More specifically, we ask whether there can exist an efficient classical algorithm that can sample from a probability distribution $p_\text{cl}(\vec{x})$ such that 
\begin{equation}
	\frac{1}{1 + \eps}p(\vec{x}) \leq p_\text{cl}(\vec{x}) \leq (1 + \eps)p(\vec{x}),
	\label{eq:PkU}
\end{equation}
where $\eps\geq0$ controls the accuracy.

In this section, we establish that even the most injective isoTNS (i.e., isoTNS with $\delta = 1/D$, which saturate the bound in \cref{delta_condition}), are hard to sample from within a multiplicative error.
Our conclusion is based on the simple insight that with post-selection, even highly injective isoTNS can encode post-selected quantum computations. It has been previously established in Ref.~\cite{Bremner2010} that families of quantum states that, with postselection, can encode postselected quantum computation are unlikely to be classically samplable to multiplicative precision since that would effectively imply that $\textsf{postBPP}=\textsf{postBQP}$ (or that the polynomial hierarchy collapses to the third level)~\cite{Bremner2010}, which is considered unlikely. This establishes a contrast between computation of local observables and sampling for isometric tensor networks, similar to what is known for constant depth quantum circuits~\cite{Terhal2002,Gao2017}.

We restrict ourselves to bond-dimension $D = 2$ and physical dimension $d = D^4 = 16$. We first note that a two-qubit unitary $U$ on the ancillary qubits can be applied, on post-selection of the physical qudit, by choosing the isoTNS tensor $P^k$ via
\begin{equation}
	P^{k}_U = \frac{1}{4}U \sigma_k \text{ for } k \in \{0, 1, 2 \dots 15\},
	\label{eq:PkSWAP}
\end{equation}
where $\sigma_0, \sigma_1 \dots \sigma_{15}$ are 2-qubit Pauli operators and $\sigma_0 = \1$.
Clearly, on postselecting the physical qudit in $\ket{0}$, we effectively apply the unitary $U$ on the ancillas. It can also be verified by explicit computation that $\norm{P^{-1}}_\infty = 2$, thus saturating the bound in \cref{delta_condition}. Furthermore, at the end of the computation, we also need to swap ancilla qubits with physical qudits.
This can be implemented in an isoTNS by using $P^k$ as
\begin{equation}
	P^k =\frac{1}{2} \ket{k_0 k_1}\!\bra{k_2 k_3},
\end{equation}
where $k := \{k_0, k_1, k_2, k_3\} \in \{0, 1\}^4$. This tensor saturates the bound in \cref{delta_condition} and thus is maximally injective. Furthermore,  post-selecting the physical qudit on $\ket{0} \cong \ket{0,0,0,0}$ effectively post-selects the ancillas on $\ket{0,0}$ and post-selecting the physical qudit on a state of the form $\ket{k_0, k_2, k_0, k_3}$ (or $\ket{k_0, k_1, k_2, k_1}$) performs a $Z-$basis measurement on one of the ancillas. Using the two maximally injective tensors shown above, it is clear that any post-selected quantum computation can be encoded into a maximally injective isoTNS with post-selection, and hence this isoTNS remains hard to sample from classically to multiplicative precision.

\subsection{Hardness transition of sampling}
\label{sec:MIPT}

In this section we take the isoTNS into which we have embedded a quantum circuit \cref{eq:BQP-circuit} and perturb it such that it becomes injective, but we choose the perturbation in such a way that it lets us prove that sampling becomes easy once the state is injective enough.
The existence of the easy phase relies on a percolation argument similar to those in measurement-induced phase transitions~\cite{Skinner2019}.

To obtain an injective version of \cref{eq:BQP-circuit}, we first need to embed each physical qubit in a larger Hilbert space $d=D^4=16$.
We decompose the physical Hilbert space at each site into four qubits.
Note that all tensors in \cref{eq:BQP-circuit} can directly be adapted to this setting for example by taking $V\to V\otimes\ket0\otimes\ket0$.

Now consider the following isometry that maps $\C^2\to (\C^2)^{\otimes 3}$ (i.e., one ancilla qubit to the ancilla qubit and two physical qubits)
\begin{equation}
	\begin{aligned}
		W &= \sum_{i,j}\ket{ij}K_{ij}=
    	\begin{array}{c}
			\begin{tikzpicture}[scale=.35,thick,rotate=90,yscale=-1,decoration={markings, mark=at position 0.8 with {\arrow{>}}}]
				\draw (0,.2)--(1,.2);
				\draw (0,.7)--(1,.7);
				\draw[color=ancilla] (-1,-.6)--(1,-.6);
				\filldraw[fill=tensor] (-1/2,-1) rectangle (1/2,1);
				\draw (0,0) node {$W$};
			\end{tikzpicture}
    	\end{array},\\
		K_{00} &=\sqrt{1/2-\delta^2}\1+i\delta\ket1\bra1 \qquad &K_{01}=i\delta\ket1\bra0,\\
		K_{11} &=\sqrt{1/2-\delta^2}\1+i\delta\ket0\bra0 \qquad &K_{10}=i\delta\ket0\bra1.
	\end{aligned}
	\label{eq:jump_ops}
\end{equation}
Measuring the physical qubits in the computational basis after applying $W$ on a single-qubit state yields unequal outcomes (01 or 10) with probability $p_{01}+p_{10}=2\delta^2$, independent of the input state.
Observing 01 or 10 projects the qubit into $\ket1$ or $\ket0$, respectively.

We now construct the one-parameter family of tensors that lead to a complexity transition in sampling, which is directly connected to what has been dubbed ``measurement-induced phase transition'' in monitored quantum circuits~\cite{Skinner2019}.
Specifically, we take the same circuit as previously, \cref{eq:BQP-circuit}, but with the perturbed gate tensor
\begin{equation}
	 {\tilde V} = ( W\otimes W) U = 
    \begin{array}{c}
		\begin{tikzpicture}[scale=.35,thick,rotate=90,yscale=-1,decoration={markings, mark=at position 0.8 with {\arrow{>}}}]
			\draw[color=ancilla] (-1, -.7)--(2.5, -.7);
			\draw[color=ancilla] (-1, 0.7)--(2.5, 0.7);
			\filldraw[fill=tensor] (-1/2,-1) rectangle (1/2,1);
			\draw (0, 0) node {$U$};
			\begin{scope}[shift={(1.5,1.25)}]
				\draw (0,.2)--(1,.2);
				\draw (0,.7)--(1,.7);
				\filldraw[fill=tensor] (-1/2,-1) rectangle (1/2,1);
				\draw (0,0) node {$W$};
			\end{scope}
			\begin{scope}[shift={(1.5,-1.25)},yscale=-1]
				\draw (0,.2)--(1,.2);
				\draw (0,.7)--(1,.7);
				\filldraw[fill=tensor] (-1/2,-1) rectangle (1/2,1);
				\draw (0,0) node {$W$};
			\end{scope}
		\end{tikzpicture}
    \end{array}.
	\label{eq:Vtilde}
\end{equation}
The other tensors we perturb to injectivity as before.
Due to our construction of $W$, now all PEPS tensors $P$ have minimum singular value $\delta$.

To simulate sampling, we evolve the ancillas as before in the direction of time, but instead of tracing over the physical sites, we sample from the sites as we go.
In general, computing this time evolution is difficult (\BQP-hard).
However, when measuring the physical qubits, in each layer and for each ancilla we have a $2\delta^2$ chance of measuring either 01 or 10 and thus resetting and disentangling the corresponding ancilla qubit, which breaks the bonds to the neighbouring tensors.
For $2\delta^2>1/2$, we are above the bond percolation threshold and large clusters are exponentially suppressed. 

To rigorously assess the computational effort required to get a certain error, we again introduce a cutoff size $s_\mathrm{th}$ and discard samples in which the largest cluster exceeds this size.
The probability to find a cluster larger than $s$ anywhere in the system, and thus the fraction of samples we have to discard, is bounded by (this time with an additional factor of $N$)
\begin{equation}
	\mathrm{Prob}(\mathcal{S} \geq s) \leq Ns^2\exp(-s/\xi_0),
	\label{eq:S}
\end{equation}
where $\xi_0>0$ is again some constant.
To ensure that the fraction of rejected samples falls below $\eps$, we can take $s_\mathrm{th}=\xi_0\log(N/\eps)+\xi_0\log(\log^3(N/\eps))$.
Clusters of this size can be contracted in $\poly(N/\eps)$ computational effort, which makes this algorithm efficient.

\section{Physical properties of injective isoTNS}
\label{sec:physical-properties}
Here we establish two facts about the physical properties of strongly injective and weakly injective isoTNS. 

\textbf{Property 1.} Weakly injective isoTNS below the threshold support arbitrary range correlations and one-dimensional subregions can have a volume law in mutual information.

\textbf{Property 2.} Strongly injective isoTNS have exponentially decaying correlations.

Property 1 directly follows from our ability to embed fault-tolerant quantum computation in the virtual space of the isoTNS.
If the fault-tolerant computation implements a random quantum circuit on the logical qubits, and we consider the one-dimensional subregion that is the output of the circuit, then the logical state of this region is close to a volume-law entangled state and thus also has a volume-law mutual information.
Similarly, we can instead choose a circuit that prepares the GHZ state of the logical qubits, which has infinite correlation length.
This establishes \emph{Property 1.}

To prove property 2, let us first introduce the approximate Markov states, which obey the uniform Markov property~\cite{Hayden2004}.
\begin{definition}[Uniform Markov property \cite{Brandao2019}]
	A state $\psi$ defined on a graph with sites $\Lambda$ satisfies the $\eps(l)$ uniform approximate Markov property if for any tripartition into regions $A$, $B$, $C$, such that $A\cup B\cup C= \Lambda$ and $\mathrm{dist}(i,j)\geq l$ for any $i\in A$ and $j\in C$, we have 
	\begin{equation}
		I(A:C|B)\leq \eps(l),
		\label{eq:markov}
	\end{equation}
	where
	\begin{equation}
		I(A:C|B)=S(\rho_{AB})+S(\rho_{BC})-S(\rho_B)-S(\rho_{ABC})
		\label{eq:mutual-information}
	\end{equation}
	and for any region $R$ and its complement $\bar R$, we define $\rho_R=\tr_{\bar R}(\rho)$.
	\label{def:markov}
\end{definition}
We show that strongly injective isoTNS satisfy the uniform Markov property with exponential decay $\eps(l)$.
For simplicity, we fix a rectangular geometry as shown in \cref{fig:ABC}.
\begin{figure}[htb]
	\begin{tikzpicture}[scale=.8,thick]
		\filldraw[fill=white] (-4,-1) rectangle (-1,1);
		\draw (-2.5,0) node {$A$};
		\filldraw[fill=white] (4,1) rectangle (1,-1);
		\draw (2.5,0) node {$C$};
		\filldraw[fill=white] (-1,-1) rectangle (1,1);
		\draw[dashed] (0,-1.1)--(0,1.1);
		\draw (-.5,0) node {$B_L$};
		\draw (.5,0) node {$B_R$};
		\draw [decorate,decoration={brace,amplitude=5pt,mirror,raise=1ex}]
  		(-1,-1) -- (1,-1) node[midway,yshift=-2em]{$l$};
		\draw [decorate,decoration={brace,amplitude=5pt,mirror,raise=1ex}]
  		(4,-1) -- (4,1) node[midway,xshift=2em]{$W$};
	\end{tikzpicture}
	\caption{For concreteness, we analyse an isoTNS with rectangular geometry with subsystems $A$, $B$, $C$. The screening region $B$ has width $W$ and length $l$ and is divided in two equally sized parts, a left half $B_L$ and right half $B_R$.}
	\label{fig:ABC}
\end{figure}

We establish
\begin{equation}
	\norm{\rho_{AC}-\rho_A\otimes\rho_C}\leq c\exp(-l/\xi),
	\label{eq:markov_property}
\end{equation}
using the same percolation trick as before. In particular, 
we introduce the state $\rho_{AC}^\gamma$, with $\gamma\subseteq B$, which is defined by taking the tensor network expression for $\ket{\psi}\bra{\psi}$ and replacing those tensors on sites in $\gamma$ through the depolarizing channel $\E_\mathrm{depol}$ and those sites in the complement $\bar\gamma=B\backslash\gamma$ through the channel $\E_1$.
Then we have 
\begin{equation}
	\rho_{AC} = \sum_\mathrm{\gamma.\gamma\subseteq B}\eta^{|\gamma|}(1-\eta)^{|B|-|\gamma|}\rho_{AC}^\gamma,
	\label{eq:rho_AC}
\end{equation}
where the sum ranges over all possible subsets $\gamma$ of $B$.
We note that each $\rho_{AC}^\gamma$ is a valid quantum state with trace one.
Percolation gives us that the vast majority of terms in \cref{eq:rho_AC} factorize if $l$ is large enough.
Specifically, \cref{eq:S} tells us that the fraction of terms with a cluster large enough to span across a distance $l$ is upper bounded by $|B|l^2\exp(-l/\xi_0)$. Since $|B|=Wl$, we only have to throw away at most a fraction of $Wl^3\exp(-l/\xi_0)$ terms to approximate \cref{eq:rho_AC} through a sum involving only factorizing density matrices.

However, this does not exclude classical correlations in $\rho_{AC}$, which also contribute to the mutual information.
These can be excluded by splitting region $B$ into a left and right region $B_L, B_R$ (see \cref{fig:ABC}). 
We then only consider those configurations $\gamma$ in \cref{eq:rho_AC} in which the configuration $\gamma_L$ in $B_L$ has at least one vertical domain wall of depolarizing channels and at the same time $\gamma_R$ also has a vertical domain wall. 
Configurations $\gamma$ in which this is not the case need to have at least one cluster of size $l/2$. 
Thus we can again bound the fraction of those configurations by $\delta=W(l/2)^3\exp(-l/(2\xi_0))$.
This allows us to approximate $\rho_{AC}$ through
\begin{equation}
	\tilde\rho_{AC} = (1-\eta)^{|B|}\sum_{\gamma_L}\frac{\eta^{|\gamma_L|}}{(1-\eta)^{|\gamma_L|}}\rho_{A}^{\gamma_L}\otimes\sum_{\gamma_R}\frac{\eta^{|\gamma_L|}}{(1-\eta)^{|\gamma_L|}}\rho_C^{\gamma_R}.
	\label{eq:rho_AC-approx}
\end{equation}
To make this approximation, we only neglected a fraction $\delta$ of terms, so $\norm{\rho_{AC}-\tilde\rho_{AC}}\leq\delta$. 
Furthermore, $\norm{\tr_C(\tilde\rho_{AC})-\rho_A}\leq\delta$ and equivalently $\norm{\tr_A(\tilde\rho_{AC})-\rho_C}\leq\delta$.
The triangle inequality thus gives us
\begin{equation}
	\norm{\rho_{AC}-\rho_A\otimes\rho_C}\leq 3\delta=3W(l/2)^3\exp(-l/(2\xi_0)).
	\label{eq:bound}
\end{equation}
We note that the correlation length $\xi_0$ only depends on the injectivity parameter, but otherwise not on the specific tensors of the state. This means that it is straightforward to apply this bound to any sequence of isoTNS of growing system size.
The factor $W$ corresponds to the width of the screening region $B$. 
\Cref{eq:bound} establishes the Markov property (\cref{def:markov}) with exponential decay and also proves uniform clustering (decay of connected correlation functions).

\section{Outlook and discussions}
We have studied the computational complexity of isoTNS, a variational class of tensor network states with the additional constraint that its tensors are isometries. 
Using a duality between the properties of these states in two dimensions and dynamics of systems in one dimension, we have shown that both computing the local expectation value and sampling from isoTNS have hard and easy phases. 
Interestingly, the hardness persists in states that are translation-invariant in the bulk (though with a disordered boundary).
This indicates that algorithms to contract infinite tensor networks based on approximating the fixed point of the transfer matrix may fail in certain instances of isoTNS.

Aside from computational complexity, we can also characterize the correlations in isoTNS based on the dynamics on the bond space.
From our circuit embeddings, it follows directly that isoTNS in the hard phase support long-range correlations and that one can find isoTNS with volume-law mutual information across selected columns (or rows) of the state. 
In contrast, we establish that in the easy phase, isoTNS obey a uniform Markov property, which in turns implies that connected correlation functions decay exponentially and that the corresponding ancilla dynamics is rapidly mixing.
Very recently, (non-injective) isoTNS with power-law correlations have been constructed~\cite{Liu2024}, where the emergence of the power law is connected to the reduced 1D dynamics of the ancillas being critical. 

IsoTNS can be defined on any graph, and one natural direction to explore in the future would be to study their complexity as a function of properties of the underlying graph. 
Another potential direction is to use the map between isoTNS and open system dynamics to define a continuous version of isoTNS.

Finally, it remains to test our results, conditions, and the algorithms we provide on isoTNS obtained from numerically optimizing for a ground state.
We note here that the algorithms we provide to compute local expectation values or to sample can be improved in a number of ways.
One promising way to improve the algorithm to compute local expectation values is to improve the decomposition of the channels. 
Instead of decomposing into depolarizing noise and the rest, one can numerically find the entanglement breaking channel with the largest rate into which the original channel can be decomposed, which would allow one to extend the easy phase.

\begin{acknowledgments}
	We acknowledge support from the Novo Nordisk Fonden under grant number NNF22OC0071934 and NNF20OC0059939. 
	DM would like to thank Freek Witteveen for insightful discussions, the International Quantum Tensor Network (IQTN) for organizing meetings that have inspired this work, and the Flatiron Institute for hosting one of them, where the idea for this paper was born.
\end{acknowledgments}

\appendix

\section{Calculation of \texorpdfstring{$\delta$}{delta} for fault tolerant construction}
\label{app:cond_num_ftc}
\emph{Perturbing gate tensor to injectivity}:
Consider the following channel on a qudit $\mathbb{C}^k$
\begin{equation}
	\E_U(\rho) = (1 - p) U \rho U^\dagger + p \tr{\rho} \frac{\1}{k}.
\end{equation}
For results concerning embedding a two-qubit unitary into an injective isoTNS, we simply set $k = 4$ and interpret $U$ as the two-qubit unitary.
$\E_U$ has $k^2$ Kraus operators $A^{\alpha, \beta}$ for $\alpha, \beta \in \{1, 2 \dots k\}$, 
\begin{align}
A^{\alpha, \beta} =\begin{cases} \gamma_0 U + \sqrt{\frac{p}{k}} \ket{\alpha}\!\bra{\alpha} U, & \text{ if }\alpha = \beta, \\
\sqrt{\frac{p}{k}}\ket{\alpha}\!\bra{\beta} U, & \text{ if }\alpha \neq \beta,
\end{cases}
\end{align}
where
\begin{equation}
	\gamma_0 = -\sqrt{\frac{p}{k}} + \sqrt{1 - p + \frac{p}{k} }.
\end{equation}
Note that the operators $A^{\alpha, \beta}$ are linearly independent, and consequently the PEPS tensor $A$ formed by the Kraus operators is injective. 
We can now compute the condition number of this tensor.
From the bond-space to physical-space, the PEPS tensor $A$ effectively implements the map
\begin{align}
   &A\ket{\alpha, \alpha} := (\1\otimes U^\dagger)\bigg[
        \gamma_0 \sum_{\alpha' = 1}^k  \ket{\alpha', \alpha'} + \sqrt{\frac{p}{k}}  \ket{\alpha, \alpha}\bigg], \\
    &A\ket{\alpha, \beta} :=  \sqrt{\frac{p}{k}}(\1\otimes U^\dagger) \ket{\alpha, \beta}.
\end{align}
Note that $\tilde{A} = (\1 \otimes U) A$ is block-diagonal within the subspaces $\mathcal{S}_\text{diag}= \text{span}\{\ket{\alpha, \alpha}: \alpha \in \{1, 2 \dots k\}\}$ and $\mathcal{S}_\text{diag}^\perp = \text{span}\{\ket{\alpha, \beta}: \alpha, \beta \in \{1, 2 \dots k\}\} $. Furthermore,
\begin{equation}
	\tilde{A}\big|_{\mathcal{S}_\text{diag}^\perp} = \sqrt{\frac{p}{k}} \1 \text{ and }{\tilde{A}}\big|_{\mathcal{S}_\text{diag}} = \sqrt{\frac{p}{k}} \1 + k \gamma_0 \ket{\Phi}\!\bra{\Phi},
\end{equation}
where $\ket{\Phi} = \sum_{\alpha = 1}^k \ket{\alpha, \alpha}/\sqrt{k}$. Thus we obtain
\begin{align}
	\delta=\sigma_\text{min}(A) = \lambda_\text{min}(\tilde{A}) = \sqrt{\frac{p}{k}}.
\end{align}

\emph{Perturbing restart tensor to injectivity}: Consider now the two qudit channel on $\mathbb{C}^k \otimes \mathbb{C}^k$
\begin{equation}
	\E_\text{res}(\rho) = (1 - p) \ket{0}\!\bra{0}\otimes \tr_1(\rho)  + p \tr(\rho) \frac{\1^{\otimes 2}}{k^2}.
\end{equation}
The $k^4$ Kraus operators for this channel are given by $A^{\alpha_1, \alpha_2; \beta}$, where $\alpha_1, \alpha_2 \in \{0, 1 \dots k - 1\}, \beta \in \{0, 1, 2 \dots k^2 - 1\}$
\begin{equation}
	A^{\alpha_1, \alpha_2; \beta} = \begin{cases}
	    \sqrt{1 - p +\frac{p}{k^2}} \ket{0}\!\bra{\alpha_2} \otimes \sigma_\beta & \text{ for }\alpha_1 = 0, \\
      \sqrt{\frac{p}{k^2}} \ket{\alpha_1}\!\bra{\alpha_2} \otimes \sigma_\beta & \text{otherwise},
	\end{cases}
\end{equation}
where $\sigma_0 = \1$, $\sigma_1, \sigma_2 \dots \sigma_{k^2 - 1}$ are traceless matrices and $\tr(\sigma_i^\dagger \sigma_j) = \delta_{i, j}$ i.e.~$\sigma_0, \sigma_1 \dots \sigma_{k^2 - 1}$ form an orthonormal basis for matrices in $\mathbb{C}^{k\times k}$. Note also that $A^{\alpha_1, \alpha_2; \beta}$ are orthogonal i.e.~$\tr(A^{\alpha_1, \alpha_2; \beta\dagger}A^{\alpha_1', \alpha_2'; \beta'}) = 0$ unless $\alpha_1 = \alpha_1', \alpha_2 = \alpha_2', \beta = \beta'$. Therefore, viewing the Kraus operators as a map from the bond indices to the physical indices, we obtain
\begin{equation}
	\sigma_\text{min}(A) = \sqrt{\frac{p}{k^2}}.
\end{equation}

\section{Bulk translation-invariant isoTNS construction}
\label{app:ti-isotns}

Given a \BQP{} decision problem, we construct an isoTNS generated by the same 2-qudit gate throughout that can decide the problem.
Recall that a \BQP{} problem can be solved by a quantum Turing machine, which we briefly recap here.
As shown in Ref.~\cite{Bernstein1997}, it is sufficient to consider a quantum Turing machine comprising a worktape, which is a semi-infinite tape of work cells, and a pointer that moves across the worktape. To solve a given decision problem $\in$ \BQP, the Turing machine is given a finite set of states $\Sigma_Q$ and a transition rule $\delta$. The set of states $\Sigma_Q$ depends on the \BQP{} problem being solved, but it is guaranteed to have an ``ACCEPT" state and a ``REJECT" state which are reached only at the end of the computation and are the answer to the \BQP{} decision problem. Both the work cells and the pointer should be considered to be quantum systems that can be entangled. The cells can be considered to be qubits. The states space of the pointer is spanned by $\ket{\sigma}\otimes \ket{x}$, where $\sigma \in \Sigma_Q$ where $\Sigma_Q$ and $x\in \mathbb{N}$ indicates where on the tape the pointer is located. The quantum Turing machine implements a unitary on the joint state of worktape and the pointer given by
\begin{align*}
&U_\text{QTM}\big(\ket{ \dots b_x \dots}\otimes \ket{\sigma, x}\big)=\nonumber\\
&\sum_{\substack{b_x'\in\{0, 1\} \\ \sigma' \in \Sigma_Q}}\bigg(\delta(b'_x, b_x, \sigma', \sigma, L)\ket{\dots b_{x}' \dots}\otimes \ket{\sigma', x - 1} +\nonumber\\
&\qquad \qquad \ \delta(b'_x, b_x, \sigma', \sigma, R) \ket{ \dots b_{x}' \dots}\otimes \ket{\sigma', x + 1} \bigg).
\end{align*}
Note that the transition rule $\delta(b', b, \sigma', \sigma, L/R)$ simply describes the complex amplitude associated with changing the state of the cell of the worktape at the pointer location from $b \to b'$, changing the Turing machine's state from $\sigma \to \sigma'$ while moving the pointer left or right. As was established in Ref.~\cite{Bernstein1997} and used in the construction in Ref.~\cite{Gottesman2009}, we can always assume that the pointer to have to move left or right (and does not need to stay at the current position). Furthermore, we can also assume that the pointer always returns to the first cell of the worktape at the end of the computation, with the pointer being in an ACCEPT and REJECT state. Furthermore, at the start of the computation, the worktape is initialized in a computational basis state which is a bit string specifying the input to the Turing machine.

\begin{figure}[htb]
    \centering
    \includegraphics[width=0.85\linewidth]{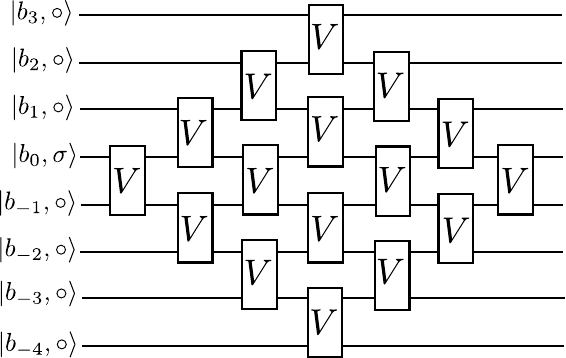}
    \caption{An oblique brickwork circuit generated by the same tensor $V$ and a translationally varying product state that effectively implements a quantum Turing machine. This circuit can equivalently be read as an isoTNS with bulk tensor with a translation-invariant bulk and a translationally varying boundary.}
    \label{fig:ti_isoTNS}
\end{figure}

We first straightforwardly encode the Turing machine into an oblique brickwork circuit formed by the same 2-qudit gate as shown in Fig.~\ref{fig:ti_isoTNS}, but with a translationally varying input state. For this, we first choose a Hilbert space for each qudit---we define a copy of $\Sigma_Q$ and distinguish it symbolically from $\Sigma_Q$ by superscripting it with a prime, $\Sigma_Q' = \{\sigma' : \sigma \in \Sigma_Q\}$ (i.e.~if $\Sigma_Q = \{\text{ACCEPT}, \text{REJECT}, \text{LOOP}, \text{CHECK}, \dots\}$ then $\Sigma_Q' = \{\text{ACCEPT}', \text{REJECT}', \text{LOOP}', \text{CHECK}' \dots\}$). Furthermore, we also introduce a filler symbol $\circ$. Each qudit in the sequential circuit has a Hilbert space spanned by 
\begin{equation}
    \ket{b, \sigma} \text{ where }b \in \{0, 1\}, \sigma \in \Sigma_Q \cup \Sigma_Q' \cup \{\circ\}.
\end{equation}
The qudits simultaneously store the worktape state and the pointer state---$b$ in $\ket{b, \sigma}$ corresponds to the date on the worktape and $\sigma$ is the state of the Turing machine. The circuit is constructed in such a way that at any given circuit step, only one of the qudit has$\sigma \neq \circ$, and this qudit corresponds to the location of the pointer. Finally, as become clear below, the purpose behind introducing the primed copy $\Sigma'_Q$ of $\Sigma_Q$ is to ensure that we can move the pointer to the left and to the right with just two qudit gates (and this redundancy would not be required if we could use 3-qudit gates). 

The two-qudit gate $V$ that is repeated throughout the circuit is given by following rules
\begin{align}
&V\ket{b_l, \sigma_l}\otimes \ket{b_r, \circ} = \sum_{{b \in \{0, 1\}}} \bigg(\nonumber\\
&\qquad \qquad \sum_{\sigma \in \Sigma_Q}\delta(b, b_l, \sigma, \sigma_l, R)\ket{b, \circ}\otimes \ket{b_r, \sigma} + \nonumber\\
&\qquad \qquad \sum_{\sigma \in \Sigma_Q'}\delta(b, b_l, \sigma', \sigma_l, L) \ket{b, \sigma'}\otimes \ket{b_r, \circ}\bigg)  \text{ if }\sigma_l \in \Sigma_Q, \nonumber\\
&V \ket{b_l, \circ}\otimes \ket{b_r, \sigma_r'}  = \ket{b_l, \sigma_r} \otimes \ket{b_r, \circ} \text{ if }\sigma' \in \Sigma_Q'.
\end{align}
On any other state, we assume that $V$ acts as identity. As depicted in Fig.~\ref{fig:ti_isoTNS} , the initial state is a product state carrying the data on the worktape and the state of the Turing machine at the center qubit. When the input to $V$ has the state $\sigma$ on the left, then $V$ effectively applies the transition rule of the quantum Turing machine ($U_\text{QTM}$)---if the pointer (the location of the state $\sigma$) is to be moved to the right, then this is done by $V$. However, $V$ cannot move the pointer to the left since it only acts on two sites---consequently, $V$ maps the state $\sigma \to \sigma'$. When the unitary $V$ applied at a later step gets an input with $\sigma'$ on the right, then it is moved to the left. With this $V$, it then follows that the oblique brickwork circuit in Fig.~\ref{fig:ti_isoTNS} effectively executes the quantum Turing machine. Since this circuit can be read as isoTNS, this shows that a general quantum computation can be embedded in an isoTNS with a translationally varying boundary, but a translation-invariant bulk tensor.

\bibliography{jabref}

\end{document}